\documentclass[aps,prb,twocolumn]{revtex4}   

\usepackage{amsmath}    
\usepackage{graphicx}   

\DeclareMathAlphabet{\mathpzc}{OT1}{pzc}{m}{it}

\begin{document}

\title{Theoretical and experimental analysis of a thin elastic cylindrical tube acting as a non-Hookean spring}

\author{Antonio \v{S}iber}
\email{asiber@ifs.hr}   
 \affiliation{Institute of physics, Bijeni\v{c}ka cesta 46, 10000 Zagreb, Croatia}
 
\author{Hrvoje Buljan}
\affiliation{Department of Physics, University of Zagreb, Bijeni\v{c}ka c. 32, 10000 Zagreb, Croatia}
\date{\today}

\begin{abstract}
We analyze the deformation and energetics of a thin elastic cylindrical tube compressed 
between two plates which are parallel to the tube axis. The deformation is studied 
theoretically using exact numerical simulations and the variational approach. These 
results are used to interpret the experimental data obtained by pressing plastic-foil 
tubes in an apparatus specially designed for this purpose. 

{\em A simplified variant of the physics that we present was used as a basis for one of the 
experimental problems posed on the 41st International Physics Olympiad held in Zagreb, Croatia in July 2010.}
\end{abstract}

\maketitle

\section{Introduction}

Springs are ubiquitous mechanical devices that can be used to absorb energy due to shock, 
vibration or exerted pressure (e.g. in vehicle suspension and clutch and brake systems, and 
some types of computer mouse devices and keyboards), to store and controlably release the energy (e.g. 
in winding clocks and children's toys) and to measure forces (e.g. in scales, spring balances, 
dynamometers and similar). Springs store the energy by changing their shape, and they are useful 
because they can deform many times in practicaly the same way, acting always the same in 
response to the applied force, before the mechanical/elastical properties of the material that they are made of 
change and the springs become worn out. Before this point, measurements on springs in their 
deformed (forced) state can be used to precisely determine the applied force - this is the basic 
principle of a scale. 

Deformation of the springs depends on the elastic properties of the material they are made of, but also 
on the construction i.e. geometry of the spring. There are many types of spring constructions, perhaps 
the best known being the helical spring. However, any piece of elastic material can be used in many different ways 
and settings so to deform in response to the applied force. The springs that are typically discussed in 
various courses of elementary physics are Hookean, i.e. their linear deformation is proportional to 
the force. For most students this becomes the hallmark of elasticity, but although the materials 
the springs are made of may be Hookean, the {\em overall} deformation of the spring can display 
all sorts of different functional dependences on the applied force, depending on the geometry 
of the spring. In particular, although a rod of a certain material may obey Hooke's law, so that its 
extension is proportional to the force stretching it, this does not mean that the cylinder made of a 
thin sheet of such a material will deform in response to the force applied perpendicularly to its axis so 
that its displacements are proportional to the force. 

In this paper, we introduce a particularly simple construction of a spring made by rolling a piece of 
thin sheet (used in strip and spiral book binding - binding cover) to form a cylindrical tube. We shall show how the  
deformation of such a spring can be predicted by applying an approximate, but completely adequate variant of 
the theory of elasticity to shells. Such an analysis enables one to construct a simple scale, but also 
to determine the elastic properties of the sheet material, as we shall demonstrate. 
The theoretical discussion of the deformation and elasticity of the spring is applied to the 
data obtained from the experimental setup that was made in order to gauge and test the spring properties.

\section{Elasticity of bending}
\label{sec:elements_theory_elasticity}

\subsection{Two-dimensional moduli of elasticity and curvatures of deformed sheets}

The elementary theory of elasticity is usually exposed on the problem of extension of rods or the 
homogeneous compression of isotropic materials. In the two cases, the elasticity parameter that 
naturally appears is the Young's modulus, that we shall denote by $E$ in the following. 

In the case of deformation of thin sheets made of elastic 
material, an important simplification arises due to the fact that the thickness of the sheet is much 
smaller from the two other characteristic dimensions (mean length and width). In such a situation, 
it is of use to introduce "two-dimensional" elasticity parameters. These 
are known under different names, perhaps the most common ones being the bending (or flexural) 
rigidity ($\kappa$) and two-dimensional Young's modulus ($Y$, sometimes also called the extensional 
stifness) \cite{Timoshenko_stability}. If the material the sheet is made of is isotropic, the two parameters are related to the 
bulk Young's modulus and Poisson ratio ($\nu$) of the material as \cite{Timoshenko_stability}
\begin{eqnarray}
Y &=& \frac{Ed}{1-\nu^2}\nonumber \\
\kappa &=& \frac{Ed^3}{12(1-\nu^2)},
\label{eq:2Dmoduli}
\end{eqnarray}
where $d$ is the thickness of the sheet. 
The above equations are derived in the advanced textbooks on the theory of elasticity by 
examining the deformation of a small element of the sheet, which are either an in-plane (in-sheet) 
stretching and shear, or an out-of-plane bending. The two types of deformation are characterized by the 
two effective moduli ($Y$ for stretching and $\kappa$ for bending). 

In general, the character of the deformation that will occur in a sheet depends on the magnitudes of the 
two elastic moduli, and the geometric constraints (which e.g. may or may not allow the bending type of 
deformation). It is typical for the sufficiently thin sheet to deform in a way 
that stores most of its energy in the bending type of deformations. In these cases, the elastic 
energy depends only on the magnitude of the bending rigidity ($\kappa$), and the in-plane stretching 
deformation that would depend on $Y$ is effectively "frozen"/forbidden (the deformation 
is {\em inextensional}). The energy of such deformations can be calculated from the curvatures 
of the sheet shape. As shown in the Appendix, for each point in a surface one can define {\em two} 
principal curvatures ($K_1$ and $K_2$) and from these, one can construct the mean ($K_M$) and gaussian ($K_G$) curvatures of the 
surface. However, in order that one surface can be transformed into some other by a pure bending (inextensional) 
type of deformation, it is necessary that their gaussian curvatures are the same. Thus, a planar 
sheet of material (with $K_G=0$) can be bent only in configurations that have a nonvanishing curvature only 
along {\em one} direction, while they need to remain flat in the perpendicular direction. Surfaces of cylinders and 
cones are examples of such configurations. In our case, the deformed spring will be a generalized cylinder, and 
the problem is thus effectively one-dimensional.

In such a case, the elastic energy of the sheet can be represented as a functional of the shape mean 
curvatures \cite{Timoshenko_stability} as 
\begin{equation}
E_{el} = \frac{\kappa}{2}\int_S K^2 dS,
\label{eq:Helfrich}
\end{equation}
where $K=2 K_M$ is twice the mean curvature (see Appendix) that in general depends on the position of the point on 
the surface $S$. For the cylindrical surfaces, the above equation can also be written as
\begin{equation}
E_{el} = \frac{\kappa h}{2} \int_{\cal C} K^2 dl, 
\label{eq:Helf2}
\end{equation}
where ${\cal C}$ is the curve outlining the shape of the cylinder base, $h$ is the cylinder height, and $dl$ is 
the inifinitesimal arc element of the curve ${\cal C}$ ($dS = h dl$). Our problem thus reduces to a planar 
problem, since the curve ${\cal C}$ with a given $h$ completely determines the shape of the deformed cylindrical spring. 

It is implicit in equations (\ref{eq:Helfrich}) and (\ref{eq:Helf2}) that the principal radii of curvatures are much 
larger than the thickness of the sheet.

\section{Experimental setup}

\begin{figure}[h]
\begin{center}
\scalebox{0.35}{\includegraphics{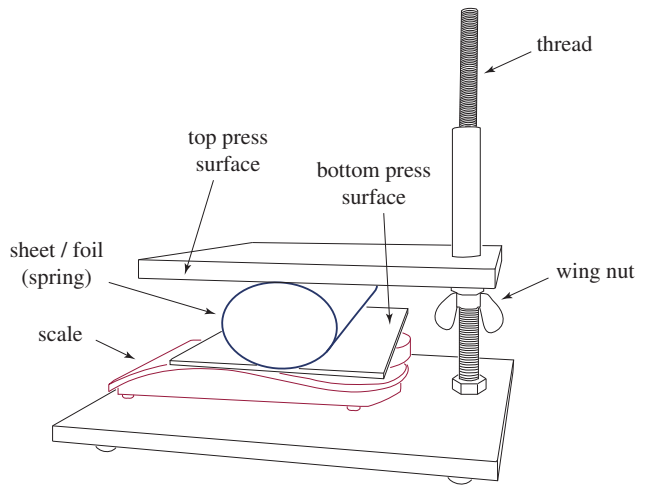}}
\caption{\label{fig:fig_apparatus}An illustration of the experimental setup.}
\end{center}
\end{figure}

An essential part of our experiment is a thin sheet made of transparent polymer 
material. The sheets that we tested are typically used as transparencies in overhead projection systems
or as transparent plastic covers for strip and spiral book binding. Their size is usually the same as those of standard paper 
sizes, A4 in our case; $W=210$ mm, $L=297$ mm. 
In order to make a spring, we roll the sheet in a cylinder, either along the width ($W$) or the length ($L$, $L>W$) of the sheet, 
and we use transparent adhesive tape to fix the cylindrical shape of the sheet. For accurate measurements, the width of 
the overlapping region of the sheet where 
the adhesive tape is applied should be as small as possible. For the sheets we tested, we have been able to reduce to width 
of overlapping region to $\sim 2$ mm. We have thus created the spring whose response we shall study by 
pressing the spring in a specially designed apparatus shown schematically in Fig. \ref{fig:fig_apparatus}. It is essentially 
a press driven by a wing nut with a known pitch which enables one to precisely determine the shift of the top press surface 
and thus the change in spring height, $2b$. As the spring is pressed, the scale 
measures the effective force that the spring exerts on it in terms of (effective) mass. The scale that we used was marketed 
as "electronic kitchen scale" (digital) with a precision of 1 g.

The process of measurement can be setup so to start from a point where the top press surface barely touches the spring - this 
can be checked by the mass reading on the scale which must remain zero. The height of the unladen spring ($2b_0$) 
should be measured at this point. As the wing nut is turned, the reading of the scale ($m_e$) should be recorded after each full 
turn of the nut. If needed, the readings can be recorded after quarter or a half turn or even after an arbitrary angle by 
mounting an angular scale above the nut which can be used to precisely determine the percentage of the turn. The smallest 
angular turns of the wing nut that we used were $\pi/3$. Since the 
pitch of the nut/thread is known, this measurement procedure (with determined $2b_0$) enables one to obtain many pairs 
of data, relating effective mass/force $m_{e}$ to the spring height $2b$.

\section{Variational approach to determination of the shape of deformed cylindrical spring}

It is of use to think of this system by turning it upside-down in a sort of gedanken experiment. One can thus recognize 
that the effective mass measured by the scale can be thought of as a mass that presses the spring from "above" due to 
gravity (the mass of the spring being neglected here). This enables 
one to pose the physical problem of spring deformation as a minimization of energy. The two types of energies in the system 
are the gravitational energy of the effective mass and the elastic (bending) energy of the deformed spring. The shape that 
shall be adopted by the spring minimizes the total energy functional, 
\begin{equation}
{\cal E} = E_p + E_{el} = 2 m_e g b + \frac{\kappa}{2} \int_S  K^2 dS.
\label{eq:total_energy}
\end{equation}

As already mentioned, the spring surface can be parametrized by the {\em profile} 
of its cross-section i.e. the closed curve ${\cal C}$ that defines it. In the language of variational calculus 
(see e.g. Ref. \onlinecite{variational_general}), of all the possible curves that have the given length (so that the spring is inextensible) and 
height of $2b$, the curve that shall actually be adopted by the spring will be the one for which the functional 
in Eq. (\ref{eq:total_energy}) is minimal. This statement can also be written as 
\begin{equation}
\frac{\delta {\cal E}}{\delta {\cal C}} = 0,
\end{equation}
where the combination $\delta / \delta {\cal C}$ symbolizes the variation of a functional ${\cal E}$ with respect to the curves ${\cal C}$. 
The problem is well 
posed in this way since the curve ${\cal C}$ can be written in a quite general fashion, so that the variation of the 
energy functional produces differential equation for the function parametrizing the curve 
(the Euler-Lagrange equations) \cite{variational_general}. However, such procedure typically results in equations that can 
be solved only numerically. Therefore, we resort to a simpler version of a variational approach that shall yield 
important analytical insights, however. 

Instead of varying the functional ${\cal E}$ over {\em all} possible functions that represent ${\cal C}$, we can 
vary it over a particular {\em subset} of all functions. The subset that is chosen typically consists of 
functions that can be studied analytically and parametrized using small number of parameters 
whose variation spans the space of subset functions. A typical example of such procedure is a variational 
determination of the ground state energy of hydrogen atom that is studied in many textbooks (see e.g. 
Ref. \onlinecite{Schiff}). In our case, we must choose the functions that are {\em (i)} closed, {\em (ii)} produce 
a given circumference of the spring profile, and {\em (iii)} are confined to a space between the two press surfaces, 
so that the profile height is $2b$. There are {\em two} substantially different categories of functions that may be 
tested as variational solutions to the 
cross-section of the deformed cylindrical sheet. The functions in the first category touch the upper and lower press 
cross-sections in single points. A typical representative of such functions is an ellipse. The functions in the second 
category touch the upper and lower press cross-sections 
along the lines of certain length. A typical representative of such functions is the function made of two 
semi-circles connected by lines whose length corresponds 
to the separation of the centers of the semi-circles. In the further, we shall term this profile as the {\em stadium}, since 
the same name is used in the literature on quantum and classical chaos for the so-called billiards of 
such a shape (sometimes also the Bunimovich stadium after the researcher who studied it \cite{chaos_billiards}).

That the two types of functions really exist as solutions to the 
real problem can be checked experimentally by using our setup. Profiles quite similar to the stadium are obtained 
in case of cylinder that is laden with sufficient mass. 
The profiles from the first category appear for quite small loads (effective masses), and it may be somewhat difficult 
to judge whether the rolled sheet touches the press surfaces only tangentially or along a region of finite area. In any case, 
one should remember that the profiles are variational functions and they thus only mimic the true solution. For example, the 
stadium profile may mimic a solution whose curvature is very low in the regions where the profile almost touches the 
press surfaces and it becomes large in the two regions where the profile separates from the press surface. Such a 
profile may in fact touch the press surfaces (that become lines in the plane of the cross-section) only in four points, yet the stadium approximation should still be an 
adequate variational try (or {\em ansatz} as it is sometimes called in the professional literature). One should keep in mind, however, 
that, since the subset of variational functions is a restricted one, the energy that is obtained is an upper limit for the exact 
energy of the system, as is always the case in a variational approach \cite{variational_general,Schiff}.

In the following two subsections we shall solve the variational problem for the two categories of spring cross-sections.

\subsection{Stadium profile}

The elastic energy of the stadium can be represented analytically by evaluating Eq. (\ref{eq:Helfrich}) for the 
profile. Flat pieces of the profile contribute nothing to the energy and the energy of the curved parts is easily calculated since 
these are two halves of a cylinder of height $h$ and radius $b$ (note that the curvature of the profile shows a discontinuity along 
the lines where the flat pieces meet the cylindrical ones). The two principal curvatures along the cylindrical 
portions are constant and given as $K_1 = b^{-1}$, $K_2 = 0$ (since the radius of curvature along the cylinder height 
is infinite). The mean curvature is $K_M = b^{-1}/2$ ($K=b^{-1}$), so that the elastic energy is 
\begin{equation}
E_{el} = \frac{\kappa}{2} \frac{2b \pi h}{b^2} = \frac{\pi \kappa h}{b}. 
\label{eq:elastic_stadium}
\end{equation}
The total energy of the system for the chosen profile is now
\begin{equation}
{\cal E} = 2 m_e g b + \frac{\pi \kappa h}{b}.
\end{equation}
Requiring that for given effective load $m_e$, the total energy be minimal leaves us with a simple variational condition on $b$, which 
is the only parameter of the profile:
\begin{equation}
\frac{\partial {\cal E}}{\partial b} = 2 m_e g - \frac{\pi \kappa h}{b^2} \equiv 0.
\end{equation}
This yields an equation for the profile characteristic radius (half of the height), 
\begin{equation}
b = \sqrt{\frac{\kappa \pi h}{2m_eg}}.
\label{eq:dep_b_m_stadium}
\end{equation}
Obviously, the above solution can be expected to be correct only for sufficiently large loads $m_e$, as it diverges as 
$m_e^{-1/2}$ when $m_e \to 0$. A more careful analysis would require that the maximally allowed value for $b$ be bounded from above by $b_0$, i.e. 
the radius of the cylinder in its unladen state - this is a form of the nonextensibility requirement to the solution. 
This yields an inequality that defines a region of the applicability of the solution, 
\begin{equation}
\sqrt{\frac{\kappa \pi h}{2m_eg}} < b_0.
\label{eq:validity_stadium}
\end{equation}
The inextensibility condition is easily satisfied whenever $b<b_0$ since the length of the flat parts of the profile can 
be adjusted as needed, without the change in the profile elastic energy.
Note, however, that the region of validity of the solution is likely to be smaller, since the stadium shape may not be the best possible 
variational {\em ansatz} throughout the region defined by Eq. (\ref{eq:validity_stadium}). 

\subsection{Elliptic profile}

The calculation of the elastic energy of an elliptic profile requires somewhat more mathematics than in the stadium case. We 
parametrize the ellipse as $x = a \cos t$, $y = b \sin t$, where $a$ and $b$ are the major and the minor axes, respectively. The 
curvature of such a profile can be calculated as explained in the Appendix, which yields $K_G=0$, as before, and 
\begin{equation}
K = \frac{ab}{(a^2 \cos^2 t + b^2 \sin^2 t)^{3/2}}.
\end{equation}
Evaluation of the integral in Eq. (\ref{eq:Helfrich}) with such a curvature yields 
\begin{equation}
E_{el} = \frac{2 \kappa h}{3 b} \left[ 2 \left( 1 + \frac{b^2}{a^2} \right) \mathpzc{E} \left( \sqrt{1 - \frac{a^2}{b^2}} \right)
- \mathpzc{K} \left( \sqrt{ 1 - \frac{a^2}{b^2} } \right) \right], 
\label{eq:elipsa_energija}
\end{equation}
where $\mathpzc{K}$ and $\mathpzc{E}$ are complete elliptic integrals of the first and second kinds, respectively \cite{notation_el_int}. The major and 
minor axes of the ellipse are connected via the condition of inextensibility of the sheet, so that 
the profile circumference always equals the circumference in the unladen state, when the radius of the cylinder is $b_0$. Equating the 
circumference of the ellipse with that of a circle (in the unladen state) yields
\begin{equation}
\mathpzc{E} \left( \sqrt{ 1 - \frac{a^2}{b^2} } \right) = \frac{b_0 \pi}{2 b}.
\label{eq:elipsa_opseg}
\end{equation}
The problem now reduces to varying Eq. (\ref{eq:elipsa_energija}), and at the same time requiring Eq. (\ref{eq:elipsa_opseg}) 
to hold. Instead of dealing with such a problem and all the inconveniences stemming from the appearance of special functions, 
we shall use the physical insight which tells us that elliptic profiles are likely to be observed for small deformations, 
i.e. when $a \approx b$. In that case, we can use the Taylor expansions of the elliptic integrals\cite{Wolfram} to relate the 
major and minor axes as $a^2 = 4b b_0 - 3b^2$, and obtain the elastic energy as
\begin{equation}
\lim _{a \to b} E_{el} = \frac{\pi h \kappa}{b} \left[ \frac{5 (b_0/b) - 4 \left( b_0 / b \right)^2 - 2}{3 - 4 ( b_0 / b ) } \right].
\label{eq:atob}
\end{equation}
The above expression differs from the elastic energy of the stadium profile [Eq. (\ref{eq:elastic_stadium})] by the multiplicative factor 
in square parentheses. Examination of this factor reveals that it is smaller than one in the interval $b \in [0.80 b_0, b_0]$, and 
it becomes larger than one when $b<0.80 b_0$. Thus, the elastic energy of the elliptic profile is smaller from the one corresponding 
to the stadium profile for initial deformation of the cylindrical spring, but for sufficiently large deformations ($b<0.80 b_0$), the 
stadium profile becomes more favorable energywise. Already at this point, we may suspect that the response of the spring will show two 
characteristic regimes separated at point where $b \approx 0.8 b_0$.

The characteristic profile parameter $b$ can be again obtained by varying the total energy which, with Eq. (\ref{eq:atob}), yields a 
polynomial equation of fitfh degree for $b$. However, since we are interested in the solutions of this equation only for small 
deformations, we may again use Taylor expansion, but this time around the point where $b = b_0$, which then yields
\begin{equation}
b = b_0 - \frac{m_e g b_0^3}{7 \pi h \kappa}.
\end{equation}

\section{Numerical solution to the problem: Universal energetics of the spring deformation}
\label{sec:numerical}

The problem of interest to us can be solved numerically. This can be done in many different ways. Perhaps the simplest one, and 
the one that we applied, is to discretize the profile of the deformed spring in $N$ points and to reformulate the energy 
functional so that it becomes a function of the coordinates of these points. Such a function can then be minimized using various numerical algorithms 
intended for such purpose. We shall use a particular variant of the conjugate gradient minimization that was successfully applied 
previously in Refs. \onlinecite{siber_nano_el,siber_cones,Siber_vir1,Siber_vir2}. The constraints of the inextensibility of the 
sheet and the impenetrability of the top and bottom press surfaces can be implemented by suplementing the elastic energy with an 
energy penalty for all configurations that violate the constraints. Such a method is common in numerical optimization with 
constraints (see e.g. Ref. \onlinecite{Siber_vir1}). All these details are not really essential for our purposes, and 
we only need to know that the highly reliable numerical solution of our problem can be obtained.

The solution can be scaled so that it becomes universal i.e. applicable to appropriately scaled measurements of springs, 
irrespective of their equilibrium radii, $b_0$, heights $h$, and bending rigidities, $\kappa$. Such a scaled solution 
depends only on an adimensional parametrization of the shape, and the energy scale appears only as a multiplicative factor, scaling 
the universal solution to the concrete case with given spring dimensions and bending rigidity. 
The adimensional parameter that uniquely determines the spring shape is 
$b/b_0$, and an appropriate scale of elastic energy is $\pi \kappa h / b_0$ (one could also use $\kappa h / b$, but it is 
more convenient to have a fixed scale of energy that does not change during the spring deformation). The energy-shape 
dependence can thus be written as
\begin{equation}
\frac{b_0}{\pi \kappa h} E_{el} = {\cal U} \left( \frac{b}{b_0} \right) \equiv \overline{E}_{el},
\end{equation}
where ${\cal U} \left( b / b_0 \right)$ is the universal function characteristic for our problem [note in particular 
that Eqs. (\ref{eq:elastic_stadium}) and (\ref{eq:elipsa_energija}) are of this form]. The appropriately scaled 
energy (adimensional) is denoted by an overline ($\overline{E}_{el}$), as will be all the adimensional 
quantitites in the following.

\begin{figure}[h]
\begin{center}
\scalebox{0.39}{\includegraphics{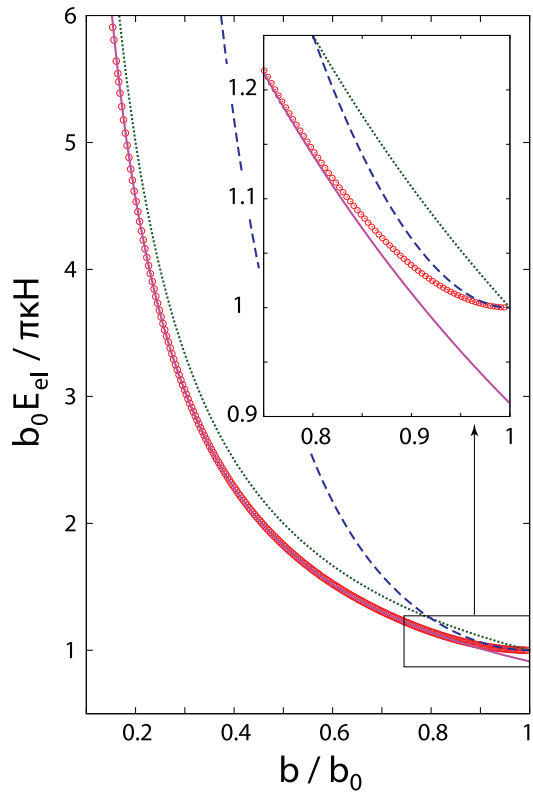}}
\caption{\label{fig:fig_energy}Theoretical predictions for the spring energies. The circles show the numerical 
results. The dotted line is the prediction of the variational method based on the stadium profile, Eq. (\ref{eq:elastic_stadium}). 
The dashed line is the variational prediction for the elliptical profile in the limit of small eccentricities, Eq. (\ref{eq:atob}). 
The full line shows the scaled variational prediction of the stadium profile, obtained by a multiplication of 
the variational results (dotted line) by a factor of 0.912.
}
\end{center}
\end{figure}

In Fig. \ref{fig:fig_energy} we show the theoretical predictions for the spring energy. The circles show the numerical 
data. The dotted line is the prediction of the variational method based on the stadium profile, Eq. (\ref{eq:elastic_stadium}). 
The dashed line is the variational prediction for the elliptical profile in the limit of small eccentricities, Eq. (\ref{eq:atob}). We 
see that the variational prediction based on the stadium profile quite nicely follows the trend of the numerical (exact) 
results in the range when $b/b_0 > 0.7$, as expected from the discussion in the previous section. The variational energy is, however, 
always {\em above} the exact results, as is always the case in variational approach. However, a simple scaling of the 
energies obtained from stadium variational results by a factor of 0.912 gives a full line that fits the numerical data to a precision 
better than 0.8 \% in the range $0.15< b/b_0 < 0.7$. One can immediately note the power of variational approach - although it may 
overestimate energies (by about 9 \% in our case), it gives strong clues regarding functional behavior of energies that are 
not always easy to interpret solely from the numerical results. The stadium variational prediction becomes worse as $b>0.7 b_0$, but 
the energies based on the elliptical profile are also quite unreliable in this interval, except quite close to the 
point where $b/b_0 = 1$. This was to be expected, having in mind the approximations that were used in deriving the 
variational prediction in Eq. (\ref{eq:atob}), the assumption of small eccentricities in particular.

\begin{figure}[h]
\begin{center}
\scalebox{0.35}{\includegraphics{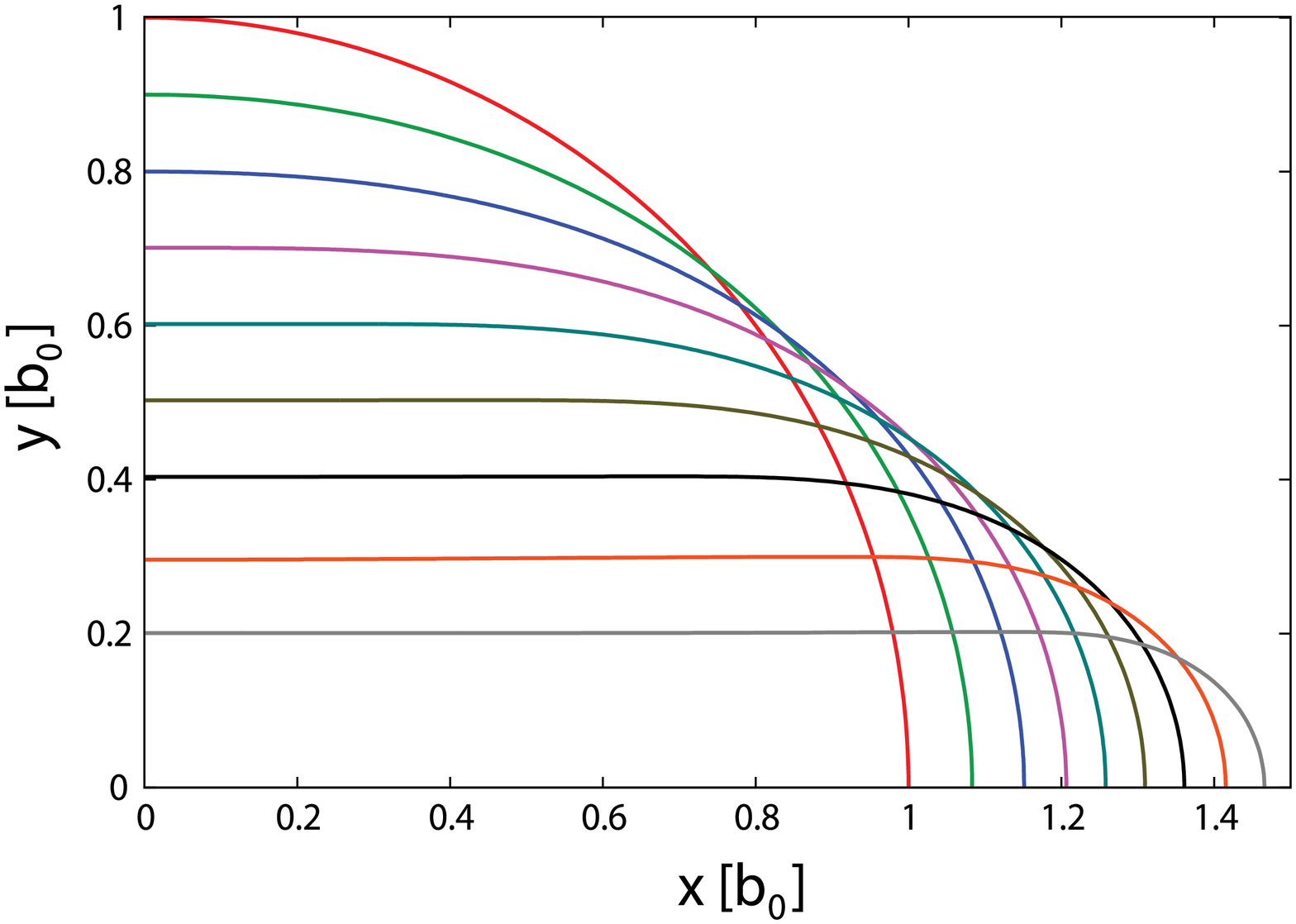}}
\caption{\label{fig:fig_profiles}The spring profiles obtained by the numerical method. The compression parameters are 
$b/b_0$ = 1.0, 0.9, 0.8, 0.7, 0.6, 0.5, 0.4, 0.3, and 0.2 (from the highest to the lowest curve).
}
\end{center}
\end{figure}

In Fig. \ref{fig:fig_profiles}, we show the profiles obtained by the numerical method. Only quarters of the profiles are shown as 
the remaining parts can be obtained by appropriate reflections about $x$ and $y$ axes. An astute reader may note a 
slight depression, concavity of the profile curve centered around $(x,y)=(0,b)$ point for sufficiently compressed 
springs ($b/b_0 \lesssim 0.3$). This feature is also easily observed in experiments that we describe in the next 
section.

\section{Experiment and theory: determining elastic properties of the sheet}

\begin{figure}[h]
\begin{center}
\scalebox{0.40}{\includegraphics{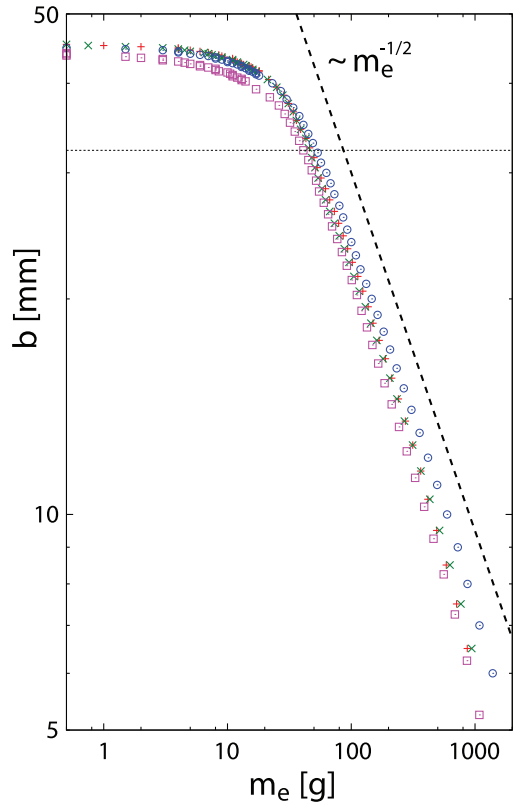}}
\caption{\label{fig:figexp1}Four representative data sets (half of the spring height, $b$ vs. scale reading, $m_e$) obtained for 
four different foils marketed and sold as "covers for binding, blue" with a thickness of 200 $\mu$m (Set 1). The foils were of A4 
dimensions (210 $\times$ 297 mm) and we rolled them along the longer side in order to make a spring, thus $h=210$ mm. The dashed line 
shows the slope expected for the $b \propto m_e^{-1/2}$ dependence in the log-log representation. The thin dotted line shows 
$b = 0.7 \langle b_0 \rangle$, where $\langle b_0 \rangle$ is the average half-height of the spring in the unladen state for the 
four sets of data.
}
\end{center}
\end{figure}

In Fig. \ref{fig:figexp1} we show four representative ''raw'' experimental data, i.e. the half of the separation 
between the two press surfaces ($b$) as a function of the mass read on the scale ($m_e$). The measurements were performed 
on four different foils, all of them belonging to the same package. The thickness of the foils is nominally 
200 $\mu$m, but our measurements of the average thickness of the foils in the package yielded 190 $\pm$ 7 $\mu$m. We shall 
denote these foils as belonging to the Set 1 in the following. We have rolled the foils along their longer side, so that $H=210$ mm. 
The separation between the two press surfaces just at the point 
where the foil barely touches the upper surface is $2 b_0 = 92$ $\pm$ 1 mm. Note that $2 b_0 \pi = 289$ mm which is 8 mm 
smaller from 297 mm (longer side of the A4 paper), and about 
half of this difference is due to small overlap of the foils that is necessary in order to apply the adhesive tape. The 
other half is due mostly to quite slight distortion of the foil under its own weight. We shall neglect this effect 
in the following, as we shall find no need for its inclusion in the data analysis.

The dashed line in the figure shows the slope expected for $b \propto m_e^{-1/2}$ dependence in the stadium 
regime, as predicted by Eq. (\ref{eq:dep_b_m_stadium}), but also by exact numerical solution shown in Fig. 
\ref{fig:fig_energy}. One can see that the predicted dependence is nicely obeyed by the data below the dotted line 
that shows $b=0.7 \langle b_0 \rangle$, where $\langle b_0 \rangle$ is the average half-height of the spring in the 
unladen state for the four sets of data. From the numerical analysis in the previous section, one finds that the 
easiest way to obtain the bending rigidity of the foils is to fit the experimental data to the 
\begin{equation}
b = \sqrt{\frac{0.912 \kappa \pi h}{2m_eg}}
\end{equation} 
dependence in the region $b < 0.7 b_0$. In addition to this, we shall perform a scaling analysis of the data, in accordance with 
the numerical results presented in Sec. \ref{sec:numerical}. The scaling analysis provides a universal description of 
the spring response, for all magnitudes of deformation and regardless of the spring bending rigidity, its height, 
and its unladen radius. It is thus 
of interest to experimentally investigate the predicted universality by studying differently shaped springs, and springs 
of different bending rigidities. It is easiest to analyze the scaling with respect to $h$, as the same sheet can be 
rolled either along its length/longer side (so that $h=W$), or along its width (so that $h=L$). Concerning the scaling with 
$\kappa$, it is not necessary that the springs be of different materials, as $\kappa$ scales with 
the sheet thickness as in Eq. (\ref{eq:2Dmoduli}), so that the sheets of the same material, but with different thicknesses 
are adequate in that respect. We have tested two additional sets of sheets, one of them sold again as binding 
covers (A4 format), but of smaller thickness (nominally 150 $\mu$m, but we measured 146 $\pm$ 8 $\mu$m). We denote the set of these foils by Set 2. Finally, the thickest set of 
sheets that we tested (A4 format, nominal thickness of 400 $\mu$m, but we measured 412 $\pm$ 4 $\mu$m) is marketed as ''flexible plastic film'' and its intended purpose is to be 
used for cutting shapes out of it (we bought it in a hobby store). The set of these foils is denoted by Set 3.

\begin{figure}[h]
\begin{center}
\scalebox{0.35}{\includegraphics{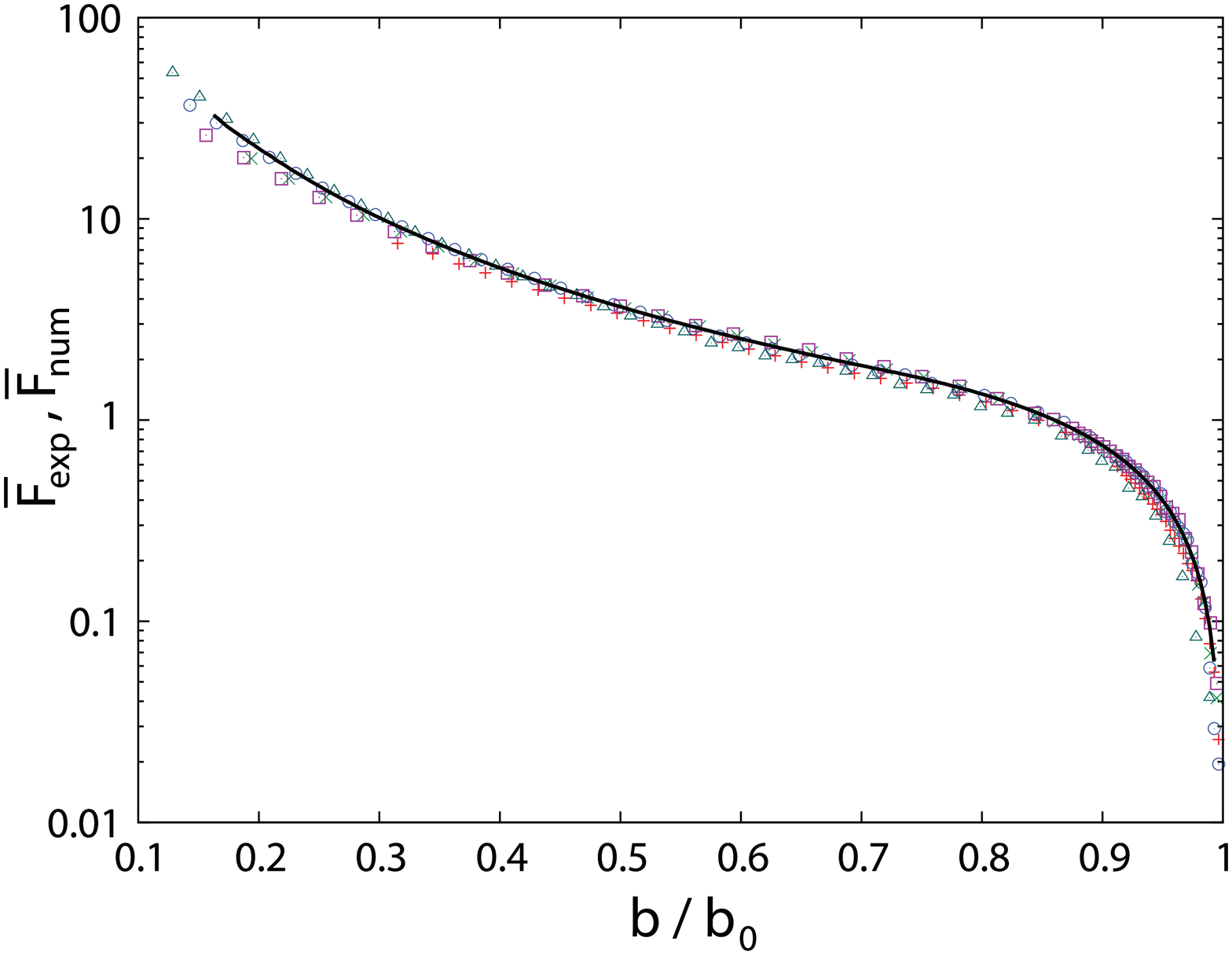}}
\caption{\label{fig:fig_scaling}Scaled forces as specified by Eqs. (\ref{eq:scaled_theo_f}) (line) and 
(\ref{eq:scaled_exp_f}) (symbols). Circles and x-es are two representative data sets corresponding 
to the two sheets from Set 1 rolled along the longer and shorter sides, respectively. Triangles and 
squares are two representative data sets corresponding to the two sheets from Set 2 rolled along 
the longer and shorter sides, respectively. Pluses show a representative data set obtained from 
a representative measurement on a sheet from Set 3 rolled along its longer side. The full line shows 
the numerical results.
}
\end{center}
\end{figure}

For all the measurements we performed, we scaled the mass readings, so to produce the adimensional experimental force $\overline{F}_{exp}$ as 
\begin{equation}
\overline{F}_{exp} = \frac{2 m_e g b_0^2}{\pi \kappa h}.
\label{eq:scaled_exp_f}
\end{equation}
The scale of force can be derived from the scale of energy, $\pi \kappa h / b_0$, simply by dividing it by the scale of 
length, $b_0$ in our case. The quantity in equation (\ref{eq:scaled_exp_f}) can be directly compared to its counterpart obtained 
from the numerical analysis, 
\begin{equation}
\overline{F}_{num} = - \frac{d \overline{E}_{el}}{d ( b/b_0 )}.
\label{eq:scaled_theo_f}
\end{equation}
Note that the factor of $2$ in Eq. (\ref{eq:scaled_exp_f}) arises from the fact that compression of the spring 
where $b$ changes by $\Delta b$ requires applying the force of $m_e g$ on a distance of $2 \Delta b$ - the same factor of $2$ is 
present in Eq. (\ref{eq:total_energy}). The comparison of the scaled experimental readings with the numerical results 
is shown in Fig. \ref{fig:fig_scaling}. One can see that the scaling predicted by the numerical results is evident 
in the experimental data through an interval of almost four order of magnitude of the force ($y$-axis; in our experimental setup, this 
corresponds to effective masses from about a gram to several kilograms). Note also that experimental data confirm the 
numerical predictions through the whole interval of deformation, in the regime where the profile can be described as a 
stadium, but also in the regime of small deformations, where the stadium ansatz fails.

The summary of the bending rigidities obtained for the sheets shown in Fig. \ref{fig:fig_scaling} is shown in Table 
\ref{tab:summary}. The fifth column of data contains the bulk Young modulus ($E$) of the sheets obtained from 
Eq. (\ref{eq:2Dmoduli}), using a value of $\kappa$ determined in the experiments and Poisson ratio of $\nu = 0.3$ that is 
typical for most materials. The bulk Young moduli are indeed in the range expected for polymeric materials such as 
nylon, for example ($E \sim$ 2-4 GPa). One should note, though, that this analysis is approximate as we did not 
measure the Poisson ratio of the sheets.

\begin{table}[h]
\begin{center}
\begin{tabular}{|c|c|c|c|c|}
\hline
Sheet & $h$ [cm] & $d$ [$\mu$m] & $\kappa$ [mJ] & $E$ [GPa] \\
\hline
Set 1 & 29.7 & 190 & 1.58 & 2.52 \\
\hline
Set 1 & 21.0 & 190 & 1.59 & 2.53 \\
\hline
Set 2 & 29.7 & 146 & 0.87 & 3.05 \\
\hline
Set 2 & 21.0 & 146 & 0.71 & 2.48 \\
\hline
Set 3 & 21.0 & 412 & 13.2 & 2.06  \\
\hline
\end{tabular}
\caption{\label{tab:summary}Summary of the properties of the sheets we used in experiments shown in 
Fig. \ref{fig:fig_scaling}. The fourth and fifth columns of data were calculated as explained in the 
text.}
\end{center}
\end{table}

This completes our theoretical and experimental analysis of the response of a thin cylindrical tube used as a spring. 

\section{Energy, elasticity, and deformation of thin shells in nano- and bio-systems}
\label{sec:applications}

Sheet-like materials and shells made of them are not uncommon at the micro- and nano-scale. These structures are often 
theoretically studied using simplified variants of theory of elasticity \cite{elasticity_bio_nano}, some of which are 
close to the one we presented. Here we mention several examples of recent research that can be understood in the 
context of physics that we presented in this work.

\subsection{Graphene, fullerenes, carbon nanotubes, graphene cones, ...}

An example of considerable recent interest is 
graphene - a single layer of carbon atoms in honeycomb arrangement \cite{graphene_rev}. It is interesting that other single-shell 
carbon structures, such as fullerenes and carbon nanotubes \cite{carbon_nanotube_rev}, can also be thought of as nanoscopic pieces of graphene material, 
cut-out from an infinite graphene plane in certain way and rolled and "glued together" \cite{tersoff_1,carbon_nanotube_rev}. 
With more elaborate cutting 
patterns, one can construct more complicated structures made of graphene, such as closed carbon cages, including 
fullerenes \cite{siber_nano_el} and carbon/graphene cones \cite{siber_cones}. Interestingly, it seems that one can calculate 
the energy of the structure, with respect to the energy of a planar piece of graphene, by accounting only for bending elastic energy 
of the shape in combination with the energy required to form {\em pentagons} in fullerenes, closed carbon nanotubes or 
carbon cones \cite{tersoff_1,siber_nano_el,siber_cones}. This is due to the fact that the shapes of these structures are locally 
pieces of cones and cylinders, so that the assumption of inextensibility holds ($K_G=0$). It is interesting that completely classical 
theory of elasticity can be succesfully applied to shapes 
in nano-domain (diameters of cylinders and shells $\sim$ 2 nm), but one should keep in mind that the bending rigidity parameter of graphene is determined by quantum physics - 
to calculate $\kappa$ for graphene, one should in principle account for change in energies of electrons in graphene that 
occurs when the planar piece of graphene is bent.

Not only the equilibrium shape and energy, but also deformation of nanotubes and fullerenes, can be studied using the theory of elasticity. 
For example in Ref. \onlinecite{cnt_hydrostatic}, the deformation of carbon nanotubes under hydrostatic pressure is studied using a 
variational approach similar to ours, but with more complicated and versatile profiles. This is a problem 
somewhat different from the one we presented, nevertheless, shapes of deformed cylinders/nanotubes quite 
similar to ours (Fig. \ref{fig:fig_profiles}) were obtained.

\subsection{Viruses and microtubules}

Biological systems abound with different structures whose characteristic dimensions are $\sim$ 10 nm. These are typically the so-called 
protein quaternary structures in which many proteins arrange in precise ways to form a larger shape. The quaternary 
structures often look like cylinders and shells of other geometries. Some typical examples are 
microtubules and protein coatings of certain viruses, such as tobacco mosaic 
virus. These structures are hollow cylinders made of many identical proteins. A different type of symmetry is more 
common in case of viruses - protein coatings of large number of viruses 
look like more or less spherical shells. The symmetry of such structures is 
quite similar to those of the icosahedral fullerenes, and analogous to pentagons and hexagons in fullerenes, the protein coatings of 
viruses are made of clusters of five and six proteins. 

Energetics and elastic response of viruses and microtubules has been studied in the literature using various 
simplifications of the theory of elasticity \cite{LMN,Siber_vir1,Siber_vir2,microtubule1}, some of which are similar to the one we presented. The experimental 
studies of such structures are typically performed using atomic force microscope (AFM) to press these structures against 
the substrate, which is an approach quite similar in spirit to the one we described. Such measurements yield information on the elastic response that can sometimes be difficult to interpret. 
In particular, when such experiments are used to measure the response of protein cylinders (tobacco mosaic virus or microtubule), 
the deformation depends also on the size of the AFM tip \cite{microtubule1,virus_AFM}. In such case, the energy of stretching may become of 
importance in determining the overall deformation. In some cases, the shells of interest cannot be treated as thin.

\subsection{Vesicles}

Vesicles are closed structures of different geometries formed by a sheet-like material that is usually a molecular bilayer. The typical 
diameters of vesicles are on the scale of micrometers \cite{Seifert97,Helfrich_original}, and due to this fact, some consider them to represent a model 
system for cells membranes. In some cases, these structures can also be investigated by the AFM technique, 
especially when they are coated with proteins. Such problems are often solved variationally, using methods similar to those we 
used in our problem \cite{clathrin_vesicles}.

\begin{acknowledgments}

We thank Hrvoje Mesi\'{c} for the ideas regarding the design of the experimental setup, that is, for 
suggesting to use a simple kitchen scale below the press for measuring the force. We thank Tomislav 
Vuleti\'{c} for designing the setup that we actually used. 

This work was supported in part by Ministry of Science, Education and Sports of Republic of Croatia 
(projects 035-0352828-2837 and 119-0000000-1015).
\end{acknowledgments}

\appendix*
\section{Curvatures of plane curves and surfaces}

Differential geometry is a part of mathematics that deals with curvatures of plane and space curves and surfaces. For our 
purposes, we shall introduce only the most elementary notions, that are sufficient to mathematically support the physics 
of our problem. Somewhat more general views on surface curvatures are also summarized in Ref. \onlinecite{Seifert97}.

\subsection{Curvature of surfaces}

\begin{figure}[h]
\begin{center}
\scalebox{0.35}{\includegraphics{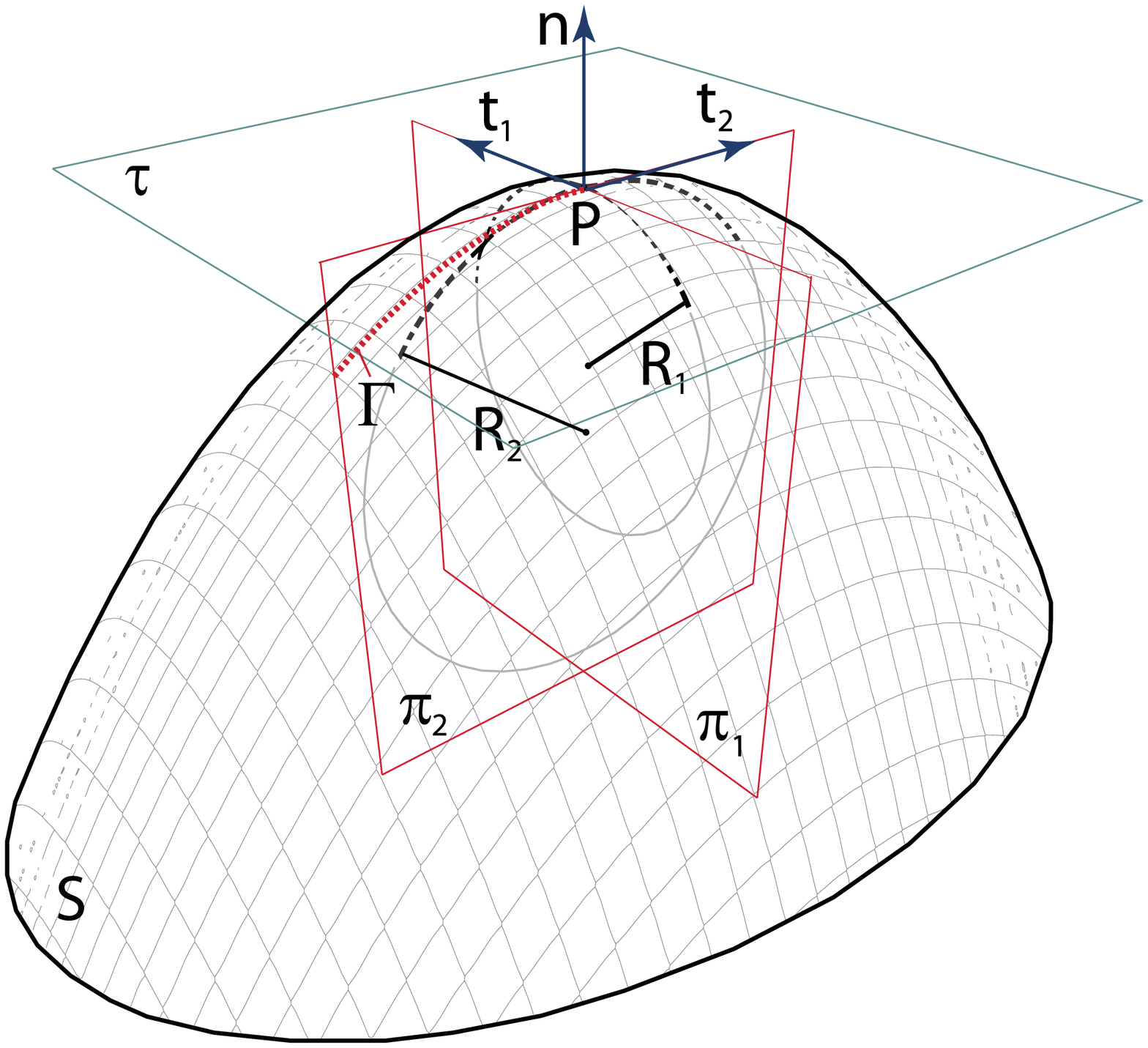}}
\caption{\label{fig:fig_curvature}A sketch illustrating concepts relating to curvature of surfaces. 
Normal vector at point $P$ of the surface $S$ is denoted by ${\bf n}$, the tangential plane at $P$ is 
denoted by $\tau$, two tangential vectors in the two principal directions are denoted by 
${\bf t}_1$ and ${\bf t}_2$, and the principal radii of curvature by $R_1$ and $R_2$. The two principal 
osculating circles are indicated by thick dashed lines. Curve $\Gamma$ 
is obtained as intersection of the surface $S$ and the plane $\pi_2$ that contains normal vector ${\bf n}$ and 
tangential direction vector ${\bf t}_2$.
}
\end{center}
\end{figure}

Here we briefly introduce the basic concepts needed to define the curvature of surfaces. These are illustrated in Fig. \ref{fig:fig_curvature}. 
The normal vector at point $P$ of the surface $S$ is ${\bf n}$. The tangent plane $\tau$ at $P$ contains all the tangent vectors  
${\bf t}$ perpendicular to ${\bf n}$. The intersection of a plane $\pi$ that contains ${\bf n}$ and a particular tangent 
vector ${\bf t}$ with a surface $S$ is a certain curve, $\Gamma$. This curve can be approximated by a circle around point $P$, 
so that {\em (i)} the circle passes through point $P$, {\em (ii)} the circle and the curve $\Gamma$ have a common tangent line at $P$,  
and {\em (iii)} the distance between the points the circle and on curve $\Gamma$ in the normal direction ${\bf n}$ 
decays as the cube of a higher power of the distance of these points to $P$ in the tangential direction ${\bf t}$. Such a 
circle is called {\em the osculating circle} of a curve $\Gamma$ (in direction ${\bf t}$). The minimum ($R_1$) and maximum ($R_2$) 
radii of all possible osculating circles at point $P$ (in all possible tangential directions) are called {\em the principal radii of curvature}. 
The two (principal) osculating circles belong to mutually perpendicular planes. The principal curvatures at point $P$ are given as
$K_1 \equiv R_1^{-1}$ and $K_2 \equiv R_2^{-1}$.

The mean ($K_M$) and gaussian ($K_G$) curvatures at point $P$ are defined as 
\begin{eqnarray}
K_M &=& \frac{K_1 + K_2}{2} \nonumber \\
K_G &=& K_1 K_2
\end{eqnarray}
We define
\begin{equation}
K \equiv 2 K_M = K_1 + K_2,
\end{equation}
i.e. twice the mean curvature.

\subsection{Curvature of plane curves}

The curvature of a plane curve given in a parametric form, $x=x(t)$, $y=y(t)$ can be obtained as
\begin{equation}
K_1 = \frac{x'y'' - y'x''}{\left( x'^2 + y'^2 \right)^{3/2}}, 
\end{equation}
where $x' \equiv dx / dt$, $y' \equiv dy / dt$.


\begin{thebibliography}{99}
\bibitem{notation_el_int} There are different notational conventions regarding the elliptic integrals. In our case, 
$\mathpzc{E}(k) \equiv \int_0^{1} (1-t^2)^{-1/2} (1 - k^2 t^2)^{1/2} dt$ and $\mathpzc{K}(k) \equiv \int_0^{1} [(1-t^2)(1 - k^2 t^2)]^{-1/2} dt$.
\end{thebibliography}
\end{document}